\documentclass[12pt]{article}
\usepackage{latexsym}
\usepackage{amsmath,amsbsy,amssymb}

\newcommand{\dd}{\mbox{\rm d}}
\newcommand{\wg}{\wedge}
\newcommand{\gam}{\gamma}
\newcommand{\Gam}{\Gamma}
\newcommand{\dg}{\dagger}

\newcommand{\tl}{\tilde}
\newcommand{\ul}{\underline}

\newcommand{\sx}{\stackrel{\bullet}{X}}

\newcommand{\DD}{\mbox{\rm D}}

\newcommand{\p}{\partial}

\newcommand{\be}{\begin{equation}}
\newcommand{\bear}{\begin{eqnarray}}
\newcommand{\ear}{\end{eqnarray}}
\newcommand{\ee}{\end{equation}}
\newcommand{\lbl}{\label}
\newcommand{\bi}{\bibitem}
\newcommand{\ci}{\cite}

\newcommand{\vs}{\vspace}

\begin{document}

\

\baselineskip .7cm 

\vs{15mm}

\begin{center}

{\LARGE \bf  Beyond the Relativistic Point Particle:
A Reciprocally Invariant System and its Generalisation
}

\vs{3mm}

Matej Pav\v si\v c

Jo\v zef Stefan Institute, Jamova 39,
1000 Ljubljana, Slovenia

e-mail: matej.pavsic@ijs.si

\vs{6mm}

{\bf Abstract}
\end{center}

\baselineskip .5cm 

{\small

We investigate a reciprocally invariant system proposed by
Low and Govaerts et al., whose action contains
both the orthogonal and the symplectic forms and is invariant
under global $O(2,4)\cap Sp(2,4)$ transformations. We find that the general
solution to the classical equations of motion has no linear term in the
evolution parameter, $\tau$, but only the oscillatory terms, and
therefore cannot represent a particle propagating in spacetime.
As a remedy, we consider a generalisation of the action by adopting a procedure
similar to that of Bars et al., who introduced the concept of a
$\tau$ derivative that is covariant under local Sp(2) transformations
between the phase space variables $x^\mu(\tau)$ and $p^\mu (\tau)$.
This system, in particular, is similar to a rigid particle whose action
contains the extrinsic curvature of the world line, which turns out to be
helical in spacetime.
Another possible generalisation is the introduction of a symplectic
potential proposed by Montesinos. We show how the latter approach
is related to Kaluza-Klein theories and
to the concept of Clifford space, a manifold whose tangent space
at any point is Clifford algebra Cl(8), a promising framework for
the unification of particles and forces.
}

Key words:  Born's reciprocity, phase space, two times, symplectic potential,
Kaluza-Klein theory, Clifford space

PACS numbers:  11.25Hf,  11.101.Lm,  04.50.Cd

\section{Introduction}

\normalsize 

\baselineskip .59cm 
 
Born's reciprocity \ci{Born} in phase space has undergone a revival in 
studies of "reciprocal relativity" \ci{Low}. The latter theory was
recently formulated in terms of an action \ci{Govaerts-Low} that contains
a term of the form ${\dot x}^\mu {\dot x}_\mu + {\dot p}^\mu {\dot p}_\mu $
and also a term of the form $p_\mu {\dot x}_\mu - {\dot p}_\mu x^\mu$. 
The dot denotes the derivative with respect to the parameter $\tau$,
which increases monotonically along a particle's world line. 
That action is invariant under global transformations of the group
${\rm O}(2,6) \cap Sp(2,6) \sim U(1,3)$. We examine
the equations of motion for such a system and find that they do not
describe a propagating particle in spacetime. This is due to the fact that the
general solution is a closed time-like or space-like loop in spacetime,
and thus represents a sort of a classical "instanton".

On the other hand, another research direction has focused on
generalising the phase space action for a relativistic point
particle \ci{Bars}. The system so obtained, besides being Lorentz-invariant, is invariant under local Sp(2) transformations that "rotate"
$x^\mu$ and $p^\mu$ into each other, and also under reparametrisations
of the world line parameter $\tau$.

In this paper, we unify the action of Govaerts et al. \ci{Govaerts-Low}
and that of Bars et al.\,\ci{Bars} in such a way that they become special
cases of
a single action. We study the corresponding equations of motion and their
general solutions. They turn out to describe, in particular, a variant of
the so called "rigid particle \ci{rigid}", i.e., a particle whose action contains the
world line's extrinsic curvature.

Finally, we employ the concept of symplectic potential \ci{SymplecticPotential}
and show that the resulting generalised action is related on one hand
to Kaluza-Klein theories, and on the other hand to Clifford algebras
and Clifford spaces\,\ci{CastroClifford,PavsicClifford,PavsicUnific},
which have turned out to
be promising for the unification of elementary particles and
interactions\,\ci{PavsicUnific}--\ci{CastroCl8}.

\section{A phase space action that contains the orthogonal and
the symplectic form }

Let us first consider\footnote{We use units in which all the
constants in front of the terms are equal to 1. Indices are raised and
lowered by the Minkowsky metric, $\eta_{\mu \nu}$}
the action proposed by Govaerts et al.\,\ci{Govaerts-Low}:
\be
I = \int \dd \tau \left [ {\dot x}^\mu {\dot x}_\mu + {\dot p}^\mu {\dot p}_\mu
+ ({\dot \theta} - \frac{1}{2} {\dot x}^\mu p_\mu + \frac{1}{2} x^\mu 
{\dot p}_\mu)^2 \right ]^{1/2},
\lbl{2.1}
\ee
where, besides the variables $x^\mu(\tau), ~\mu = 0,1,2,3$, denoting the
position of the particle in spacetime, there are also additional
variables, $p^\mu$ and $\theta$. In order to elucidate their roles,
let us consider the equations of motion derived from the action (\ref{2.1}):
\bear
&& \delta x^\mu \, : ~~~~  \frac{\dd}{\dd \tau} \left (
\frac{{\dot x}^\mu - \frac{1}{2} \mu \,p^\mu}{\sqrt{Q}} \right )
- \frac{1}{2} \frac{\mu \, {\dot p}^\mu}{\sqrt{Q}} = 0,  \lbl{2.2} \\
&&\delta p^\mu \, : ~~~~ \frac{\dd}{\dd \tau} \left (
\frac{{\dot p}^\mu + \frac{1}{2} \mu \, x^\mu}{\sqrt{Q}} \right )
+ \frac{1}{2} \frac{\mu \,{\dot x}^\mu}{\sqrt{Q}} = 0, \lbl{2.3} \\
&&\delta \theta \, : ~~~~ \frac{\dd}{\dd \tau}
\left ( \frac{\mu}{\sqrt{Q}} \right ) = 0,
\lbl{2.4}
\ear
where $\mu \equiv {\dot \theta} - \frac{1}{2}
{\dot x}^\mu p_\mu + \frac{1}{2} x^\mu    {\dot p}_\mu $
and $\sqrt{Q} \equiv ({\dot x}^2 + {\dot p}^2 + \mu^2)^{1/2}$.
From Eqs.(\ref{2.2})--(\ref{2.4}) we have
\bear
\frac{\dd}{\dd \tau} \left ( \frac{{\dot x}^\mu}{\sqrt{Q}} \right )
= C {\dot p}^\mu,   \lbl{2.5} \\
\frac{\dd}{\dd \tau} \left ( \frac{{\dot p}^\mu}{\sqrt{Q}} \right )
= - C {\dot x}^\mu,   \lbl{2.6}
\ear
where we have taken into account Eq.\,(\ref{2.4}), which implies
\be
\frac{\mu}{\sqrt{Q}} = C.
\lbl{2.7}
\ee

From the first order Eqs. (\ref{2.5}) and (\ref{2.6}),
for variables $x^\mu and ~p^\mu$ we obtain the second order equations for
variables $x^\mu$:
\be
\frac{1}{\sqrt{Q}} \frac{\dd}{\dd \tau} \left ( \frac{1}{\sqrt{Q}}
\frac{\dd}{\dd \tau} \left (\frac{{\dot x}^\mu}{\sqrt{Q}} \right ) \right )
+ C^2  \frac{{\dot x}^\mu}{\sqrt{Q}} = 0. \lbl{2.8} \ee
If we choose a parametrisation in which $\sqrt{Q} = 1$, then the equations of motion
(\ref{2.8}) become:
\be
\stackrel{...}{x}^\mu + C^2 {\dot x}^\mu = 0,
\lbl{2.9}
\ee
which after the integration gives
\be
{\ddot x}^\mu + C^2 x^\mu = a^\mu.
\lbl{2.10}
\ee
This is an ordinary equation for oscillatory motion whose general solution is
\be
x^\mu = A^\mu \, {\rm cos} \, \omega \tau + B^\mu \,{\rm sin} \omega \tau
+ \frac{a^\mu}{\omega^2},
\lbl{2.11}
\ee
with $\omega^2 \equiv C^2$.

A general solution is thus a closed curve (loop) in spacetime or a sort of "instanton". There is no term associated with translational motion.

Let us now consider Eq.\,(\ref{2.7}), which in the gauge $\sqrt{Q} =1$ reads
\be
{\dot \theta} - \frac{1}{2}(p_\mu {\dot x}^\mu  -– x_\mu {\dot p}^\mu )= C.
\lbl{2.12}
\ee
This gives
\be
\theta = C \tau + \int \dd \tau \,
\frac{1}{2} (p_\mu {\dot x}^\mu –- x^\mu {\dot p}_\mu ) 
=  C \tau + \oint  \frac{1}{2} \, (p_\mu \dd x^\mu –- x_\mu \dd p^\mu ),
\lbl{2.13}
\ee
where the integral in the last term runs over the loop in the space of
$\lbrace x^\mu , p^\mu \rbrace$. We see that variable $\theta$ is related
to the phase, and is equal to the phase if the integration constant, $C$, is zero.

By inspecting Eq.\,(\ref{2.5}) we find that ${\dot p}^\mu$ is acceleration.
In the gauge $\sqrt{Q} = 1$, we have ${\dot p}^\mu = C^{-1} {\ddot x}^\mu$.
However, Eqs.\,(\ref{2.2}) and (\ref{2.3}) tell us that $p_\mu$ is not the conjugate
momentum to $x^\mu$. The conjugate momentum is
\be
\Pi_\mu^x = \frac{\p L}{\p {\dot x}^\mu} = \frac{{\dot x}_\mu}{\sqrt{Q}}
- \frac{1}{2} \frac{\mu}{\sqrt{Q}} \, p_\mu,
\lbl{2.14}
\ee
whereas the integration of (\ref{2.5}) gives
\be
\frac{{\dot x}_\mu}{\sqrt{Q}} = \frac{\mu}{\sqrt{Q}}\, p_\mu + C_1.
\lbl{2.15}
\ee
Thus we have the following relation between the canonical momentum,
$\Pi_\mu^x$, and the variables, $p_\mu$:
\be
\Pi_\mu^x = \frac{1}{2} \frac{\mu}{\sqrt{Q}}\, p_\mu + C_1.
\lbl{2.16}
\ee
Equation (\ref{2.15}) says that $p^\mu$ up to an integration constant $C_1$
equal to the velocity in the subspace spanned by $\lbrace x^\mu \rbrace$.
Analogous relationships hold for the momentum
$\Pi_\mu^p = \p L/\p {\dot p}^\mu$, which is conjugate to $p_\mu$.

\section{Including local Sp(2) transformations}

The solution of Eq.\,(\ref{2.11}) does not represent a physical propagation of
a relativistic particle in spacetime $\lbrace x^\mu \rbrace$ since a
term of the form $v^\mu \tau $ is missing\footnote{Evolution of the form
(\ref{2.11}) could correspond to a
motion of relativistic particle in AdS space\,\ci{PlyushchayPrivate}.
In that case
translations have another form.
For the discussion of the nonrelativistic limit of such a system see
\ci{Plyushchay1}.}. 
In order to include such a
term, we must generalise the action in Eq.\,(\ref{2.1}) using the procedure found in ref.\,\ci{Bars}.

Let us first rewrite the action in Eq.\,(\ref{2.1}) into a more compact notation:
\be
I = \int \dd \tau \left [ {\dot z}^a G_{ab} {\dot z}^b + ({\dot \theta}
- \frac{1}{2} {\dot z}^a J_{ab} z^b )^2 \right ]^{1/2},
\lbl{3.1}
\ee
where $z^a \equiv (x^\mu,p^\mu),~\mu = 0,1,2,3$, and
\be
G_{ab}  =    \begin{pmatrix}
            \eta_{\mu \nu}  &    0\\
             0    & \eta_{\mu \nu}  \\
               \end{pmatrix} \quad {\rm and} \qquad 
     J_{ab} = \begin{pmatrix}
              0    &      \eta_{\mu \nu}\\
                                   - \eta_{\mu \nu}   &   0\\
              \end{pmatrix}                                   
\lbl{3.3}
\ee
are the orthogonal and the symplectic metrics, respectively.

The action is invariant under global transformations,
\be
z \rightarrow z'' = \Gam z,
\lbl{3.4}
\ee
that preserve both the orthogonal and the symplectic form:
\bear
{\tl z'}' G z'' &=& {\tl z} G z 
\lbl{3.5}\\
{\tl z'}' J {\dot z'}' &=& {\tl z} J {\dot z}.
\lbl{3.6}
\ear
This implies:
\bear
{\tl \Gam} G \Gam &=& G
\lbl{3.7} \\
{\tl \Gam} J \Gam &=& J,
\lbl{3.8}
\ear
where the tilde denotes the transpose. The two conditions (\ref{3.7})
and (\ref{3.8})
are satisfied\,\ci{Low} by $8 \times 8$ matrices,
$\Gam \in {\rm O} (2,6) \cap {\rm Sp}(2,6)$:
\be
      \Gam = \begin{pmatrix} \Lambda  &  -M\\
                                   M       &       \Lambda\\
             \end{pmatrix}
\lbl{3.9}
\ee
such that
\be
{\tl \Lambda} \eta \Lambda + {\tl M} \eta M = \eta
\lbl{3.10}
\ee
\be
{\tl \Lambda} \eta M - {\tl M} \eta \Lambda= \eta.
\lbl{3.11}
\ee
Matrices $\Gamma$ can be written in the form:
\be
\Gam = {\bf 1} \otimes \Lambda + {\bf i} \otimes M,
\lbl{3.12}
\ee
where
\be
   {\bf 1} = \begin{pmatrix}
             1 & 0\\
             0 & 1\\
            \end{pmatrix} \quad {\rm} {\rm and} \quad 
    {\bf i} = \begin{pmatrix}
              0 & -1 \\
              1 & 0 \\
            \end{pmatrix} .
\lbl{3.13}
\ee
Here, ${\bf i}$, which satisfies ${\bf i}^2 = -1$, is a representation of the
imaginary unit, $i$.

More abstractly, $\Gam$ are thus complex $4 \times 4$ matrices such that
\be
\Gam = \Lambda + i M   \in U(1,3),
\lbl{3.14}
\ee
satisfying
\be
\Gam^\dg \eta \Gam = ({\tl \Lambda} +– i {\tl M}) \eta
(\Lambda + i M) = \eta,
\lbl{3.15}
\ee
which implies Eqs.\,(\ref{3.10}) and (\ref{3.11}).

The orthogonal and the symplectic metrics have the form
\be
G = {\bf 1} \otimes \eta ~, ~~~~~{\rm i.e.,}~~G_{ab} \equiv
G_{(i \mu)(j \nu)} = \delta_{ij} \eta_{\mu \nu}
\lbl{3.16}
\ee
\be
J = \boldsymbol{\epsilon} \otimes \eta ~, ~~~~~{\rm i.e.,}~~J_{ab} \equiv
J_{(i \mu)(j \nu)} = \epsilon_{ij} \eta_{\mu \nu},
\lbl{3.17}
\ee
where $i,j = 1,2$.

With $z^a \equiv z^{i \mu} \equiv X^{i \mu} = (X^\mu,p^\mu)$ the
action (\ref{3.1}) can be rewritten as
\be
I = \int \dd \tau \left [  {\dot X}^{i \mu} \delta_{ij} \eta_{\mu \nu}
{\dot X}^{j \nu} + ({\dot \theta} - \frac{1}{2} {\dot X}^{i \mu} \epsilon_{ij}
\eta_{\mu \nu} X^{j \nu} )^2 \right ]^{1/2}.
\lbl{3.18}
\ee
According to ref.\,\ci{Bars} we will consider $(X^{1 \mu},X^{2 \mu})
\equiv (x^\mu, p^\mu)$ as a doublet under local continuous Sp(2)
duality symmetry:
\be
     X'^{i \mu} (\tau) = {S^i}_k (\tau) X^{k \mu} (\tau) , ~~~{\rm shortly}
     \quad X'^\mu = S X^\mu.
\lbl{3.19}
\ee
The $\tau$-derivative transforms as ${\dot X}'^{i \mu} =
{S^i}_k {\dot X}^{k \mu} +{{\dot S}^i}_k X^{k \mu}$.
This can be converted into the covariant $\tau$-derivative
by introducing the Sp(2) gauge fields:
\be
     \frac{\DD X^{i \mu}}{\DD \tau} \equiv \sx^{i \mu} =
     {\dot X}^{i \mu} - {A^i}_k X^{k \mu},
\lbl{3.20}
\ee
that transform according to $A' = S A S^{-1} + {\dot S} S^{-1}$
so that $\sx '^{i \mu} (\tau) = {S^i}_k (\tau) \sx^{k \mu}$.

Under local Sp(2) we have
\be
    \sx^{i \mu} \delta_{ij} \sx^{j \nu} \eta_{\mu \nu} \rightarrow
    \sx^{k \mu} {S^i}_k \delta_{ij} {S^j}_\ell \sx^{\ell \nu} \eta_{\mu \nu},
\lbl{3.20a}
\ee
\be
    \sx^{i \mu} \epsilon_{ij} X^{j \nu} \eta_{\mu \nu}
    \rightarrow \sx^{k \mu} {S^i}_k \epsilon_{ij} {S^j}_\ell X^{\ell \nu}
    \eta_{\mu \nu} = \sx^{k \mu} \epsilon_{k \ell} X^{\ell \nu} \eta_{\mu \nu}.
\lbl{3.20b}
\ee
Since ${S^i}_k \delta_{ij} {S^j}_\ell \neq \delta_{ij}$, we have to
replace $\delta_{ij}$ with a general symmetric metric, $g_{ij}$, that
along the particle's world line depends on $\tau$ and transforms according
to $g_{ij} \rightarrow g'_{ij} = {S^k}_i g_{k \ell} {S^\ell}_j$.

Let us therefore consider the following action:
\be
    I = \int \dd \tau \left [ \sx^{i \mu} g_{ij} \eta_{\mu \nu} \sx^{j \nu}
      + ({\dot \theta} - \frac{1}{2} \sx^{i \mu} \epsilon_{ij} \eta_{\mu \nu}
      X^{j \nu})^2 \right ]^{1/2},
\lbl{3.21}
\ee
which is invariant under global transformations $\Gam \in {\rm O}(2,6)\cap
{\rm Sp}(2,6)$, and under local Sp(2) duality transformations that
interchange $x^\mu$ and $p^\mu$. We now include a constant $M$ with the
dimension of mass.

The general metric, $G_{ab}$, in the space of $z^a \equiv X^{i \mu}$ is not
``flat," and one should add a kinetic term for $G_{ab}$ to the action.
By taking a particular solution that gives a certain $G_{ab}(z)$, we can
consider it to be a background metric in which our (test) particle moves.
Once $G_{ab}(z)$ is fixed in such a way, we no longer vary it. Let us
now assume that there exists a domain in the space of $z^a$ in which
the metric can be approximated as the product $G_{ab} = g_{ij} \eta_{\mu \nu}$, where
$i,j = 1,2$ and $\mu, \nu =0,1,2,3$. Thus, we arrive at the action (\ref{3.21})
for a particle moving in such a background metric. For simplicity reasons
we will consider the case in which $g_{ij} = \delta_{ij}$. The latter
equality, of course, no longer holds if we perform a (local)
Sp(2) transformation.

Let us now consider the gauge fields, ${A^i}_k (\tau)$ considered in ref.\,
\ci{Bars}. Indices of those fields are lowered and raised by the Sp(2)
invariant metric, $\epsilon_{ij}$, and its inverse, $\epsilon^{ij}$.
So we have ${A^n}_k = A_{mk}\epsilon^{mn}$, where $A_{mk} = A_{km}$.
The three independent fields, $A_{mk}$, are associated with the three
independent Sp(2) transformations. The choice of $A_{mk}$ corresponds to the
choice of the local gauge. Variation of the action (\ref{3.21}) with respect
to $A_{rs}$ gives
\be
    - X^{(s \mu} {\epsilon^{r)}}_k \eta_{\mu \nu} \sx^{k \nu} +
    \frac{1}{2} \, \mu \, X^{r \mu} \eta_{\mu \nu} X^{s \nu} = 0,
\lbl{3.22}
\ee
where
\be
    \mu \equiv  {\dot \theta} - \frac{1}{2} \sx^{i \mu} \epsilon_{ij}
    \eta_{\mu \nu} X^{j \nu}
\lbl{3.22a}
\ee
and ${\epsilon^r}_k = \epsilon^{rm} g_{mk} = \epsilon^{rm} \delta_{mk}$,
since we take $g_{mk} = \delta_{mk}$.

The constraints in Eq. (\ref{3.22}) are due to the local Sp(2) symmetry of our
action (\ref{3.21}) and they differ from the constraints obtained
in refs.\,\ci{Bars} because our action (\ref{3.21}) is different.
However, equations of motion imply a particular case which approaches
that of ref.\,\ci{Bars}. Namely, if $\sx^{i \mu} = {\dot X}^{i \mu}
- {A^i}_k X^{k \mu} = 0$, then eq.\,(\ref{3.22}) reduces to the
constrainst of ref.\,\ci{Bars}. 
Symmetrisation is defined according to
$a^{(r} b^{s)} \equiv \frac{1}{2}(a^r b^s + a^s b^r)$.

Variation of (\ref{3.21}) with respect to $X^{\ell \rho}$ gives
\be
  \frac{\dd}{\dd \tau} \left ( \frac{g_{\ell k} \eta_{\rho \nu}
  \sx^{k \nu}}{\sqrt{Q}} \right ) + \frac{1}{\sqrt{Q}} {A^i}_\ell
  g_{ik} \eta_{\rho \nu} \sx^{k \nu} + \frac{\mu}{\sqrt{Q}}
  \sx^{i \nu} \epsilon_{i \ell} \eta_{\nu \rho} = 0,
\lbl{3.23}
\ee
where
\be
     Q \equiv \sx^{i \mu} g_{ij} \eta_{\mu \nu} \sx^{j \nu} 
      + ({\dot \theta} - \frac{1}{2} \sx^{i \mu}
     \epsilon_{ij} \eta_{\mu \nu} X^{j \nu})^2,
\lbl{3.23a}
\ee
whereas variation with respect to $\theta$ gives Eq.\,(\ref{2.4}), in which
$Q$ and $\mu$ are now defined according to (\ref{3.23a}) and (\ref{3.22a}),
respectively.

Multiplying (\ref{3.23}) by $\eta^{\rho \sigma} g^{\ell j}$, and
taking into account\footnote{The same symbol $A$ has already been
reserved if indices are lowered and raised with $\epsilon_{ij}$ and
$\epsilon^{ij}$, respectively, therefore we now use a different symbol.}
that ${A^i}_k g_{i \ell} = B_{k \ell}$, $B_{k \ell} g^{\ell j} =
{A_k}^j = - {A^i}_k$ we have:
\be
    \frac{\dd}{\dd \tau} \left ( \frac{\sx^{j \sigma}}{\sqrt{Q}} \right )
     - {A^j}_k \frac{\sx^{k \sigma}}{\sqrt{Q}}
    + \frac{\mu}{\sqrt{Q}} \sx^{i \sigma} {\epsilon_i}^j = 0.
\lbl{3.24}
\ee
In the first two terms we recognise the covariant derivative that is analogous
to (\ref{3.20}) and, therefore, Eq.\,(\ref{3.24}) can be condensed to
\be
    \frac{\DD}{\DD \tau} \left ( \frac{\sx^{j \sigma}}{\sqrt{Q}} \right )
    + \frac{\mu}{\sqrt{Q}} \sx^{i \sigma} {\epsilon_i}^j = 0.
\lbl{3.25}
\ee
In the limit $\mu/\sqrt{Q} \gg 1$, Eq.\,(\ref{3.25}) becomes
$\sx^{i \sigma} {\epsilon_i}^j = 0$, which are the equations of
motion considered in ref.\,\ci{Bars}.

As an example, let us now choose a gauge in which ${A^1}_1 = - A_{12}
=-{A^2}_2=0$, ${A^2}_1 = A_{11} =0$, whereas ${A^1}_2 = - A_{22} \neq 0$,
and does not change with $\tau$. Then the equations of motion become
\bear
 &&\frac{\dd}{\dd \tau} \left ( \frac{\sx^{1 \sigma}}{\sqrt{Q}} \right )
  + (A_{22} - \mu) \frac{\sx^{2 \sigma}}{\sqrt{Q}} = 0, \lbl{3.26} \\
 &&\frac{\dd}{\dd \tau} \left ( \frac{\sx^{2 \sigma}}{\sqrt{Q}} \right )
  + \mu \frac{\sx^{1 \sigma}}{\sqrt{Q}} = 0.  \lbl{3.27}
\ear

Using now the notation $X^{1 \sigma} \equiv x^\mu$, $X^{2\sigma}
\equiv p^\sigma$, and the relations $\sx^{1 \sigma} = {\dot X}^{1 \sigma}
+ A_{22} X^{2 \sigma}$ and $\sx^{2 \sigma} = {\dot X}^{2 \sigma}$, that hold
in the above gauge, we have
\be
    \frac{\dd}{\dd \tau} \left ( \frac{{\dot x}^\sigma + 
    A_{22} p^\sigma}{\sqrt{Q}} \right ) + \frac{1}{\sqrt{Q}} (A_{22} - \mu)
    {\dot p}^\sigma = 0,
\lbl{3.28}
\ee
\be
  \frac{\dd}{\dd \tau} \left ( \frac{{\dot p}^\sigma}{\sqrt{Q}} \right )
  + \frac{\mu}{\sqrt{Q}} ({\dot x}^\sigma + A_{22} p^\sigma ) = 0.
\lbl{3.29}
\ee
Recall that due to the equation of motion (\ref{2.4}), $\mu/\sqrt{Q} = C$,
where $C$ is an integration constant.

Let us choose a reparametrisation of $\tau$ such that $\sqrt{Q} = 1$,
so that $\mu = C$. Equations\,(\ref{3.28}) and (\ref{3.29}) then read
\bear
     &&{\ddot x}^\mu + (2 A_{22} - C) {\dot p}^\mu = 0, \lbl{3.30} \\
     &&{\ddot p}^\mu + C {\dot x}^\mu + C A_{22} p^\mu = 0. \lbl{3.31}
\ear
Differentiating (\ref{3.31}) and inserting
\be
   {\dot p}^\mu = \frac{-{\ddot x}^\mu}{2 A_{22} - C}
\lbl{3.32}
\ee
we obtain
\be
     \stackrel{....}{x}^\mu + \omega^2 {\ddot x}^\mu = 0,
\lbl{3.33}
\ee
where $\omega^2 = C (C - A_{22})$. A general solution is
\be
   x^\mu = - \frac{a^\mu}{\omega^2} \, {\rm cos} \, \omega \tau
      - \frac{b^\mu}{\omega^2} \, {\rm sin} \, \omega \tau
      + v^\mu \tau  + x_0^\mu
\lbl{3.34}
\ee
Integrating (\ref{3.32}) gives
\be
     p^\mu = \frac{-{\dot x}^\mu}{2 A_{22} - C} + k^\mu.
\lbl{3.35}
\ee
If we insert this into the second equation of motion (\ref{3.31}), then
after some straightforward computation we obtain the following relation between
$A_{22}$ and the integration constants $v^\mu$, $k^\mu$, and $C$:
\be
      v^\mu = - C A_{22} k^\mu.
\lbl{3.36}
\ee
If $A_{22} = 0$, then the 4-velocity $v^\mu$ vanishes and we arrive at
the case described in Section 2. Non vanishing $A_{22}$ is thus necessary
in order to have a propagating particle in 4-dimensional spacetime.

If we insert the solution (\ref{3.34}),(\ref{3.35}) into the
constraints (\ref{3.22}), we obtain the following conditions on the
integration constants
\be
   a^\mu a_\mu = b^\mu b_\mu = v^\mu v_\mu = a^\mu b_\mu = a^\mu v_\mu =
   b^\mu v_\mu = 0
\lbl{3.36a}
\ee 
This can be satisfied if the spacetime has at least two time-like dimensions.

We have arrived at the solution (\ref{3.34}),(\ref{3.35}) by choosing
$Q = \sx^{i \mu} g_{ij} \eta_{\mu \nu} \sx^{j \nu} + \mu^2 = 1$
(choice of gauge). But, using (\ref{3.36a}), we find that the first
term in the latter equation vanishes. Therefore, the integration constant
$C= \mu/\sqrt{Q} = \mu$ cannot be arbitrary,  but must be equal to 1. This
should not depend on choice of gauge. Indeed, if we choose an arbitrary
gauge $Q = \sx^{i \mu} g_{ij} \eta_{\mu \nu} \sx^{j \nu} + \mu^2 = 
f(X^{i \mu})$, we still have $C=1$.
 
A special treatment is necessary, if the quadratic form $Q$, entering the
action (\ref{3.21}), is light-light, $Q=0$. Then the solution still has
the form (\ref{3.34}),(\ref{3.35}), together with the condition (\ref{3.36a}),
but with the vanishing integration constant $C=0$.

A general solution, Eq. (\ref{3.34}), is a helix. If the integration
constants $a^\mu,~b^\mu$ vanish, then the oscillating term in (\ref{3.34})
disappears, so that $x^\mu = v^\mu \tau + x_0^\mu$, which is a
rectilinear motion of the ordinary relativistic particle. This can
also be seen by putting $\sx^{i \mu} =0$ into the equations of
motion, (\ref{3.28}) and (\ref{3.29}), which then reduce to the phase space
equations for a relativistic point particle:
\be
    {\dot p}^\mu = 0 ~~~~~~~ {\rm and} ~~~~~~~~ {\dot x}^\mu + A_{22} \, p^\mu = 0
\lbl{3.37}
\ee

Here the index $\mu$ runs not only over the four dimensions of
spacetime with signature $(+ - - -)$, but also over one extra
time like dimension, and---following ref.\,\ci{Bars}---also over
one extra space like dimension.  Dimensions
and signatures occurring in eqs.\,(\ref{3.1})--(\ref{3.17}) should then
be modified accordingly. Ghosts can be eliminated
by using the Sp(2) gauge symmetry as in ref.\,\ci{Bars}, which
brings us then to an effective system with one time. In ref.\,\ci{Bars}
it is shown how various gauge choices lead to various
1-time systems with different physical interpretation. It would
be interesting to explore this for the case of the generalised action
(\ref{3.21}) -- a project which is beyond the scope of the
present letter.

The precursors of 2-time physics for a single particle were
models describing two interacting particles (or a particle and string)
\ci{BarsA}. Instead of interpreting coordinates $X^{1 \mu},~X^{2 \mu}$ in
eq.\,(\ref{3.21}) as describing a single particle in higher dimensions,
we could alternatively interpret them as describing two interacting
particles in lower dimensions. According to such interpretation,
our model would be a variant of the class of models discussed in
refs.\,\ci{BarsA}.

The theory considered here goes beyond the relativistic point particle.
The fourth order equations of motion (\ref{3.33}) can be derived
from the second order action:
\be
    I[x^\mu,{\dot x}^\mu] = \int \dd \tau ( {\ddot x}^\mu {\ddot x}_\mu
    - \omega^2 {\dot x}^\mu {\dot x}_\mu ),
\lbl{3.38}
\ee
which is a functional of variables $x^\mu (\tau)$ and velocities
${\dot x}^\mu (\tau )$. The second order derivative term in (\ref{3.38})
is a gauge fixed {\it extrinsic curvature} of the world line. Thus, we have
a variant of the ``rigid particle"\,\ci{rigid}, and the variables
$(x^\mu (\tau),\,{\dot x}^\mu (\tau ))$ correspond to the variables
$(x^\mu (\tau),~p^\mu (\tau))$ of the original action (\ref{3.21}).
The latter action leads to the relation (\ref{3.35}), which says that
$p^\mu$ are proportional to ${\dot x}^\mu$ apart from the integration constants,
$k^\mu$. Fourth order differential
equations of the form similar to eq.\,(\ref{3.33})
appear in the model of the massless particle with rigidity (curvature)\,
\ci{Plyushchay2}.

\section{Further considerations}
\subsection{Symplectic potential}

Let us now again consider the action (\ref{3.1}) and rewrite it in terms
of the symplectic potential, $\theta_a (z)$, as is suggested in
ref.\,\ci{SymplecticPotential}:
\be
    I = M \int \dd \tau \left [ {\dot z}^a G_{ab} {\dot z}^b +
        ({\dot \theta} - \theta_a {\dot z}^a )^2 \right ]^{1/2} .
\lbl{4.1}
\ee

The symplectic 2-form is defined according to
\be
   \omega_{ab} = \p_a \theta_b - \p_b \theta_a.
\lbl{4.2}
\ee
In particular, if
\be
   \theta_a = \frac{1}{2} J_{ab} z^b = \frac{1}{2} (p_\mu, -x_\mu )
\lbl{4.3}
\ee
then $\omega_{ab} = J_{ab}$. In general, $\theta_a$ need not be a field
determined by Eq.\,(\ref{4.3}), but can be an arbitrary background field.
Introduction of the vector field $\theta_a$ renders the action (\ref{4.1}) invariant
under general coordinate transformations, $z^a \rightarrow z'^a = z'^a (z)$.

The canonical variables are $(z^a, \theta)$ and the conjugate momenta are
\bear
    &&\Pi_a^z = \frac{\p L}{\p {\dot z}^a} = 
    M \, \frac{G_{ac}{\dot z}^c}{\sqrt{Q}} - \Pi^\theta \theta_a, \lbl{4.4}\\
    &&\Pi^\theta = \frac{\p L}{\p \theta} = 
    M \, \frac{{\dot \theta}- \theta_a {\dot z}^a}{\sqrt{Q}},
\lbl{4.5}
\ear
where $\sqrt{Q} \equiv \sqrt{{\dot z}^a G_{ab} {\dot z}^b +
({\dot \theta} - \theta_a z^a)^2  }$.

From the identity
\be
   \frac{1}{(\sqrt{Q})^2} \left [{\dot z}^a G_{ab} {\dot z}^b +
({\dot \theta} - \theta_a z^a)^2 \right ] - 1 = 0
\lbl{4.6}
\ee
and using (\ref{4.4}) and (\ref{4.5}), we find the following constraint amongst
the momenta:
\be
   \Phi = (\Pi_a^z + \Pi^\theta \theta_a) G^{ab} (\Pi_b^z + \Pi^\theta \theta_b)
   + (\Pi^\theta)^2 - M^2 = 0.
\lbl{4.7}
\ee
Upon quantisation this becomes a Klein-Gordon like equation:
\be
      \Phi \Psi = 0,
\lbl{4.8}
\ee
where momenta are replaced by the operators: $\Pi_a^z \rightarrow {\hat \Pi}_a^z
= - i \p/\p z^a$ and $~\Pi^\theta \rightarrow {\hat \Pi}^\theta
= - i \p/\p \theta$. 

Taking the square root, we obtain a generalised Dirac equation:
\be
   \left [ \gam^a ({\hat \Pi}_a + {\hat \Pi}_\theta \theta_a)
   + \gam_{\Pi^\theta} {\hat \Pi}^\theta + \gam_M M \right ] \Psi = 0.
\lbl{4.9}
\ee
Here $\gam^a$ are Clifford numbers, satisfying $\gam^a \cdot \gam^b \equiv
\frac{1}{2} (\gam^a \gam^b + \gam^b \gam^a) = G^{ab}$, whereas
$\gam_{\Pi^\theta}$ and $\gam_M$ are Clifford numbers that satisfy
$\gam_{\Pi^\theta}^2 =1$, $\gam_M^2=1$, $\gam_{\Pi^\theta} \cdot \gam_M = 0$,
$\gam^a \cdot \gam_{\Pi^\theta}=0$, and $\gam^a \cdot \gam_M = 0$.

Solutions to the Klein-Gordon like equation (\ref{4.8}), 
for a particular $\theta_a$ given in (\ref{4.3}), were considered
by Govaerts et al.\,\ci{Govaerts-Low}. They obtained a continuous mass
spectrum of bosonic states. By tuning the value of a constant parameter
in the action, namely the cosmological constant\footnote{Our $M^2$
corresponds to $\Lambda$ of ref.\,\ci{Govaerts-Low}.}
term $M^2$, it is possible to project out negative norm states,
whilst tachyonic states\footnote{Tachyons occur already at the classical
level since the world lines are loops in spacetime (see solution
(\ref{2.11})). }
cannot be avoided. It is beyond the scope of the present letter to go into
an investigation of the spectrum of fermionic states satisfying the
generalised Dirac equation (\ref{4.9}).
 
\subsection{Maxwell-like equations for the symplectic potential}

A next possible step is to include a kinetic term for the field
$\theta_a$. The total classical action is then
\be
  I[Z^a, \theta^a] = M \int \dd \tau \left [ {\dot Z}^a G_{ab} {\dot Z}^b +
        ({\dot \theta} - \theta_a {\dot Z}^a )^2 \right ]^{1/2}
        + \kappa \int \dd^8 z \, \omega_{ab}\,  \omega^{ab}.
\lbl{4.10}
\ee
The equations of motion for variables\footnote{We now use a capital
symbol in order to distinguish the variables $Z^a (\tau)$, which describe the motion of a
particle, from the coordinates $z^a$ of the embedding space.}
$Z^a$ are just like the Lorentz force law:
\be
   \frac{1}{\sqrt{Q}}
   \frac{\dd}{\dd \tau} \left ( \frac{G_{cb} {\dot Z}^b}{\sqrt{Q}} \right )
   +  \frac{\Pi^\theta}{M}
   \frac{\omega_{cb} {\dot Z}^b}{\sqrt{Q}} = 0.
\lbl{4.10a}
\ee
If we vary (\ref{4.10}) with respect to $\theta_c$ we obtain:
\be
   \int \dd \tau \, \Pi_\theta \, \frac{{\dot Z}_c}{\sqrt{Q}}
   \delta^8 (z - Z(\tau)) = \kappa \, \p_a {\omega^a}_c.
\lbl{4.11}
\ee
These are just like Maxwell equations with a point particle source, where
$\theta_a (z)$ corresponds to the electromagnetic potential,
${\omega^a}_c$ corresponds to the field strength, and $\Pi^\theta$ corresponds to the
electric charge. The difference is that we now have an 8-dimensional
space with coordinates $z^a$.

Eight dimensional\footnote{A modification of the spacetime dimensioinality
and signature is necessary, if we consider the action (\ref{3.21})
with the local Sp(2) gauge symmetry. The action (\ref{4.10}) does not have
such symmetry.}
 action (\ref{4.10}) and the equations of motion,
(\ref{4.10a}) and (\ref{4.11}), contain the ordinary 4-dimensional
Maxwell equations as a particular case if ${\dot z}^a = 
({\dot x}^\mu,0)$, i.e., if ${\dot p}^\mu = 0$.

Another particular case is a possible solution to (\ref{4.10a}) and (\ref{4.11})
such that $\theta_a (z)$ is given by Eq.\,(\ref{4.3}). This is expected
to be the case if the variables $z^a (\tau)$ satisfy
Eqs.\,(\ref{2.5})--(\ref{2.7}), whose solution (\ref{2.11})
is a helix\footnote{It is a loop in Minkowski space $\lbrace x^\mu \rbrace$,
but there is another dimension corresponding to $\theta(\tau)$ that
satisfies Eq.\,(\ref{2.13}).}
in the 5-dimensional space $\lbrace x^\mu, \theta \rbrace$. Since
$p^\mu (\tau)$ also satisfies an equation analogous to (\ref{2.13}),
we have in fact a helix in 9-dimensional space
$\lbrace x^\mu, p^\mu, \theta \rbrace$. Such helical motion of a point
particle can in turn be considered as a ``source" for the symplectic
potential (\ref{4.3}), considered as a solution to Eq.\,(\ref{4.11}).
And vice versa: in the presence of the potential (\ref{4.3}), the particle's
world line is a helix. However, it is important to check whether such a
self-consistent helical solution of Eqs.\,(\ref{4.10a}),(\ref{4.11})
for {\it a single particle} indeed takes place. In other words, it
remains to be explored whether the symplectic potential (\ref{4.3})
can only be due to the presence of other sources, excluding our
``test particle" moving in such a background, or perhaps it can
also be due to a self-consistent helical motion of a single particle.

However, if we take another possible interpretation, discussed at the
end of Sec.\,3, namely that the action (\ref{4.10}) describes two
interacting particles with coordinates $X^{1 \mu},~X^{2 \mu}$, which
altogether form eight coordinates $X^{i \mu} \equiv Z^a$, $i=1,2,
~\mu=0,1,2,3$, then it is clear that the trajectories are not straight
lines.

\subsection{Invoking Kaluza-Klein theory}

The form
\be
     Q = {\dot Z}^a G_{ab} {\dot Z}^b +
        ({\dot \theta} - \theta_a {\dot Z}^a)^2
\lbl{4.12}
\ee
in the action (\ref{4.10}) is known in Kaluza-Klein theories.
Instead of four dimensions, $x^\mu$ plus an extra dimension, $x^5$, we
now have eight dimensions, $z^a$ plus an extra dimension, $\theta$.
The potential, $\theta_a$, is proportional to a mixed component of the
metric tensor in 9-dimensions, analogous to $g_{\mu 5}$. Thus,
$Q$ is just a 9-dimensional ``quadratic" form:
\be
    Q = {\dot Z}^A q_{AB} {\dot Z}^B,
\lbl{4.13}
\ee
where $Z^A = (Z^a, \theta)$ and $q_{AB}$ is a generic metric in
9-dimensions. We arrive at the form (\ref{4.12}) by taking the
well known Kaluza-Klein ansatz\,\ci{Kaluza-Klein} for metric
$q_{AB}$.

We see that the point-particle action, based on the simple quadratic
form $Q$ in 9-dimensions, contains the action (\ref{2.1}) of
Govaerts et al.\,\ci{Govaerts-Low} as a particular case.

\subsection{Invoking Clifford space}

Instead of considering $\theta$ as an extra dimension, we can consider it as
a scalar coordinate of the {\it Clifford space}. The concept of the
Clifford space, $C$, has been investigated in refs.\,\ci{CastroClifford,
PavsicClifford} (See also \ci{Pezzaglia}). This is a manifold whose
tangent space at any point
of $C$ is Clifford algebra. Now, instead of starting from
the 4-dimensional Minkowski space, we can start from the 8-dimensional
space whose points are described by coordinates $z^a = (x^\mu, p^\mu)$,
and build up the corresponding 256-dimensional Clifford space ($2^8=256$).
The basis of its tangent space is given by
\be
  \gam^A = \lbrace {\ul 1}, \gam^a, \gam^{a_1}\wg\gam^{a_2},...,
  \gam^{a_1}\wg \gam^{a_2} \wg ... \wg \gam^{a_8} \rbrace.
\lbl{4.14}
\ee
While $z^a$ are coordinates associated with vectors
$\gam^a$, the extra quantity, $\theta$, can be considered as a
coordinate associated
with the scalar unit ${\ul 1}$. We now take $z^A$ as coordinates,
$q_{AB}$ as the metric tensor, and $Q$ of Eq.\,(\ref{4.13}) as the
quadratic form of the full 256-dimensional space, $C$. A generic
quadratic form is given by expression (\ref{4.13}).
As a special case, for a suitable choice of metric $q_{AB}$,
we obtain the Kaluza-Klein like quadratic form (\ref{4.12}).
The latter form, in turn, for a particular choice of $\theta_a$
given in Eq.\,(\ref{4.3}),
becomes the form entering the action (\ref{3.1}) of
Govaerts et al.\,\ci{Govaerts-Low}.

\section{Conclusion}

Born's reciprocity principle between coordinates and momenta has led
investigators to the theory of Govaerts et al.\,\ci{Govaerts-Low}
on the one hand, and to the theory of Bars et al.\,\ci{Bars}
on the other hand. The former theory is based
on an action that contains an orthogonal and a symplectic form,
whilst the latter theory contains a symplectic form
only, the action being invariant under local Sp(2) transformations,
besides being Lorentz invariant. We have shown how these two theories
can be unified by means of a single action (\ref{3.21}). The
constraints (\ref{3.22})
that arise from varying the action with respect to the Sp(2) gauge
fields, $A_{ij}$, contain, besides the term considered by Bars et al.\,\ci{Bars},
an addition term. We have found that the generalised constraints also
require at least two time like dimensions, just as in refs.\,\ci{Bars}.
Their important result that in 2-time physics ghosts can be eliminated
by using the
Sp(2) gauge symmetry holds for this generalised case as well. What remains
to be explored is how various gauge choices---in analogy to the
results of refs.\,\ci{Bars}---lead to various 1-time systems with
different physical interpretations.

One possible generalisation of the action by Govaerts
et al.\,\ci{Govaerts-Low} is to put it into a form
invariant under local Sp(2) transformations that transform
$x^\mu$ and $p^\mu$ into each other, as discussed above.
Another possible generalisation
is in introducing the symplectic potential, $\theta_a$. By doing so,
we obtain an action that is invariant under general coordinate
transformations of coordinates $z^a = (x^\mu, p^\mu)$. Such action
can be considered, \`a la Kaluza-Klein, as coming from an action
that is just a line element in 9-dimensions, $\theta$ being
the ninth dimension---besides the eight dimensions, $z^a$;
the symplectic potential is then related to certain components
of the 9-dimensional metric. Moreover, we can go even further
by invoking the concept of Clifford space, $C$, and consider $\theta$
as one of the dimensions of $C$. Instead of a 16-dimensional
Clifford space whose tangent space at any point is Clifford algebra Cl(4),
we have now a 256-dimensional Clifford space with Cl(8) as a tangent
space. Unification of fundamental interaction within the framework of
Clifford algebras has been investigated by many authors
\ci{ClifUnific,CastroCl8,PavsicUnific},
and there exist strong arguments that Cl(8) is more suitable for such
purpose than Cl(4). However, if starting from 8-dimensional vector
space, a question arises as to what is a physical meaning of the
extra four dimensions. In this paper we have pointed out that those
"extra dimensions" come from phase space: a particle is described not
only by its position, $x^\mu$, in 4-dimensional spacetime, but also by its
momentum, $p^\mu$. That physics has to be formulated in phase space
has been proposed in a number of works\,\ci{Low,CastroBorn,PhaseSpace}.
In this letter we have looked at such concepts from a broader perspective,
found connections amongst various directions of research
known from the literature, and pointed out how they are related to the
unification of particles and forces within the framework of $Cl(8)$.

\vs{5mm}

\centerline{Acknowledgement}

This work was supported by the Ministry of High Education,
Science, and Technology of Slovenia.

\end{document}